\documentstyle[12pt]{article}
\newcommand{\th}{\vartheta}
\newcommand{\ph}{\varphi}
\newcommand{\uns}[3]{S_{\hat{#1}\hat{#2}\hat{#3}}}
\newcommand{\unq}[3]{Q_{\hat{#1}\hat{#2}\hat{#3}}}
\newcommand{\una}[3]{A^{\hat{#1}\hat{#2}}{}_{#3}}
\newcommand{\unantia}[3]{A^{[\hat{#1}\hat{#2}]}{}_{#3}}
\newcommand{\unsyma}[3]{A^{(\hat{#1}\hat{#2})}{}_{#3}}
\newcommand{\unaa}[3]{A^{\hat{#1}\hat{#2}}{}_{\hat{#3}}}
\newcommand{\unantiaa}[3]{A^{[\hat{#1}\hat{#2}]}{}_{\hat{#3}}}
\newcommand{\unsymaa}[3]{A^{(\hat{#1}\hat{#2})}{}_{\hat{#3}}}
\newcommand{\tthreefour}[2]{t_{\hat{#1}}{}^{\hat{#2}}}
\newcommand{\jss}[3]{\ifnum #1=#2 {J_{\hat{#1}\hat{#2}}{}^{
\hat{#3}}}\else{J_{\scriptstyle (\hat{#1}\hat{#2})}{}^{
\hat{#3}}}\fi}
\newcommand{\jaa}[3]{J_{\scriptstyle [\hat{#1}\hat{#2}]}{}^{%
\hat{#3}}}
\newcommand{\f}[4]{F^{#1#2}{}_{#3#4}}
\newcommand{\christ}{{\renewcommand{\baselinestretch}{1.0}
\large\normalsize$
\displaystyle \left\{\begin{array}{l}
\lambda \\ \mu\,\nu \end{array}\right\}$}}

\newcommand{\Bi}{f^{\mbox{\scriptsize I}}}
\newcommand{\Bii}{f^{\mbox{\scriptsize II}}}
\newcommand{\Biii}{f^{\mbox{\scriptsize III}}}
\begin{document}
\title{Isotropic cosmology in metric-affine
gauge theory of gravity}
\author{A.V. Minkevich and  A.S. Garkun\footnotemark[2]}
\renewcommand{\thefootnote}{\fnsymbol{footnote}}
\date{}
\maketitle
\begin{abstract}
Geometrical structure of homogeneous isotropic models in the
frame of the metric-affine gauge theory of gravity (MAGT) is
analyzed. By using general form of gravitational Lagrangian
including both a scalar curvature and various invariants
quadratic in the curvature, torsion and nonmetricity tensors,
gravitational equations of MAGT for homogeneous isotropic models
are deduced. It is shown, that obtained gravitational equations
lead to generalized cosmological Friedmann equation for the
metrics by certain restrictions on indefinite parameters of
gravitational Lagrangian. Isotropic models in the Weyl-cartan
space-time are discussed.\\
\\
{\raggedright  arch-ive/9805007}
\end{abstract}
\section{Introduction}
\footnotetext[2]{Address: Department of Theoretical Physics,
Belarussian State University,
av. F.Skorina 4, 220050 Minsk, Republic of Belarus\\
E-mail: mink@phys.bsu.unibel.by }
\renewcommand{\thefootnote}{\arabic{footnote}}
\setcounter{footnote}{0}
The application of gauge approach to gravity leads to
generalization of einsteinian general relativity theory (GR)
\cite{ma1}. There are different gauge theories of gravity (GT) in
dependence on the choice of gravitational gauge group and
gravitational Lagrangian: Poincare gauge theory (PGT),
metric-affine gauge theory (MAGT), its simplest variant --- gauge
theory in the Weyl-Cartan space-time (WCGT) et al. Gauge theories
of gravity were applied to resolve the problem of cosmological
singularity of GR (see \cite{ma2} and refs. given here). Regular
isotropic cosmological models were built in the frame of PGT,
MAGT (WCGT) by using sufficiently general gravitational
Lagrangian $L_G$ including both a scalar curvature and terms
quadratic in the curvature and torsion tensors. Although the
number and structure of gravitational equations of PGT, MAGT,
WCGT are essentially different, as it was shown these GT lead to
identical cosmological equations for the metrics in the case of
homogeneous isotropic models \cite{ma3,ma4}. Their investigation
allowed to make the following general conclusion. Satisfying the
correspondence principle with GR in the case of gravitating
systems with rather small energy densities, GT lead to
essentially different physical consequences in comparison with GR
in the case of gravitating systems at extreme conditions with
extremely high energy densities and pressures, the dynamics of
which depends significantly on terms of $L_G$ quadratic in the
curvature tensor.

In the present paper further study of homogeneous isotropic
models in the frame of MAGT (WCGT) is given by using more general
gravitational Lagrangian including also terms quadratic in the
nonmetricity tensor. In Sec. 2 geometrical structure of
homogeneous isotropic space in MAGT is discussed. In Sec. 3
gravitational Lagrangian and gravitational equations of MAGT
(WCGT) are given. In Sec. 4 gravitational equations of MAGT
(WCGT) for homogeneous isotropic models are deduced. In Sec. 5
the generalized cosmological Friedmann equation is introduced and
certain simplest isotropic models are discussed.

\section{Geometrical structure of ho\-mo\-ge\-neous iso\-tro\-pic
mo\-dels in MAGT}
Geometrical structure of space-time in the frame of MAGT is
defined by three tensors\footnote{$\mu$, $\nu$, \ldots are
holonomic (world) indices; $i$, $k$ \ldots are anholonomic
(tetrad) indices. Numerical tetrad indices are denoted by means
of a sign "\^{}" over them.}: metrics $g_{\mu\nu}$, torsion
$S^\lambda{}_{\mu\nu}=\Gamma^\lambda{}_{[\mu\nu ]}$
($\Gamma^\lambda{}_{\mu\nu}$ is space-time connection), nonmetricity
$Q_{\mu\nu\lambda}={\stackrel{
\scriptstyle\Gamma}{\nabla}_{\lambda}%
g_{\mu\nu}}$ ($\stackrel{\Gamma}{\nabla}$ is the covariant
derivative
defined by $\Gamma^{\lambda}{}_{\mu\nu}$ ). The connection
$\Gamma^{\lambda}{}_{\mu\nu}$ can be expressed by means of the
Christoffel symbols \christ, torsion and nonmetricity as follows
\begin{equation}
\label{conn}
\Gamma^{\lambda}{}_{\mu\nu}={\displaystyle \left\{\begin{array}{l}
\lambda \\ \mu\,\nu \end{array}\right\}}+S^{\lambda}{}_{\mu\nu}+
S_{\mu\nu}{}^{\lambda}
+S_{\nu\mu}{}^{\lambda}+\frac{1}{2}\left(Q_{\mu\nu}{}^{\lambda}-
Q_{\mu}{}^{\lambda}{}_{\nu}-Q_{\nu}{}^{\lambda}{}_{\mu}\right)
\end{equation}
Let us consider the form of tensors $S^{\lambda}{}_{\mu\nu}$ and
$Q_{\mu\nu\lambda}$ in the case of homogeneous isotropic models
in the comoving reference frame by using spherical coordinates
$\left(\,x^1=r,\:x^2=\th,\: x^3=\ph\,\right)$ and metrics in the
form of the Robertson-Walker metrics
\begin{equation}\label{rwmetric}
g_{\mu\nu}=diag\left(\;1,\; -\frac{\displaystyle R^2\left(t\right)}%
{\displaystyle 1-kr^2},\;-R^2\left(t\right)r^2,
\;-R^2\left(t\right)r^2\sin^2\th\,\right),
\end{equation}
where $R(t)$ is a scale factor, $k=-1,0,+1$ in case of open, flat
and close models respectively. The torsion tensor
$S^\lambda{}_{\mu\nu}$ is determined by two functions of time
$S(t)$ and $\tilde{S}(t)$, which define the following
nonvanishing components \cite{ma3}
\begin{equation}
\begin{array}{l}
S^1{}_{10}=S^2{}_{20}=S^3{}_{30}=S\left(t\right),\\
S_{123}=S_{231}=S_{312}=\tilde{S}\left(t\right)\frac{
\displaystyle R^3r^2}%
{\displaystyle \sqrt{1-kr^2}}\sin\th.
\end{array}
\end{equation}
By supposing that the theory is invariant under space inversions
we have $\tilde{S}(t)=0$ and the torsion tensor is defined by the
only function $S(t)$. The nonmetricity tensor is defined by three
functions of time $Q_i(t)$ ($i=1,\:2,\:3$) \cite{ma5}, so we have
the following nonvanishing components:
\begin{equation}
\begin{array}{c}
Q_{000}=Q_2(t),\,Q_{110}=\frac{\displaystyle Q_1
R^2}{\displaystyle 1-kr^2},\, Q_{220}=Q_1 R^2r^2,\\
Q_{330}=Q_1R^2r^2\sin^2\th,\, Q_{011}=\frac{\displaystyle
Q_3R^2}{\displaystyle 1-kr^2},\\ Q_{022}=Q_3R^2r^2,\,
Q_{033}=Q_3R^2r^2\sin^2\th .
\end{array}
\end{equation}
By choosing the tetrad corresponding to the metrics
(\ref{rwmetric}) in the diagonal form
\begin{equation}\label{rwtetr}
\begin{array}{c}
h^i{}_{\mu}=diag\left(1,\ \frac{R\left(t\right)}{\sqrt{1-kr^2}},
\ R\left(t\right)r,\ R\left(t\right)r\sin\th\right),\\
\left(g_{\mu\nu}=\eta_{ik}h^i{}_{\mu}h^k{}_{\nu},\ \eta_{ik}=
diag\left(1,\,-1,\,-1,\,-1\right)\right),
\end{array}
\end{equation}
we can write nonvanishing anholonomic components of torsion and
nonmetricity as
\begin{equation}\label{tor_nonm}
\begin{array}{l}
\uns{1}{0}{1}=\uns{2}{0}{2}=\uns{3}{0}{3}=S\left(t\right),\\
\unq{1}{1}{0}=\unq{2}{2}{0}=\unq{3}{3}{0}=Q_1\left(t\right),\\
\unq{0}{0}{0}=Q_2\left(t\right),\\
\unq{0}{1}{1}=\unq{0}{2}{2}=\unq{0}{3}{3}=Q_3\left(t\right).
\end{array} \end{equation}
Thus the geometrical structure of homogeneous isotropic models
studied below is determined by five functions of time: $R(t)$,
$S(t)$, $Q_i(t)$ ($i=1,2,3$).

Using (\ref{conn}) -- (\ref{tor_nonm}) one finds nonvanishing
components of gauge potentials $A^{ik}{}_\mu$ as follows
\begin{equation}\label{aconnec}
A^{ik}{}_{\mu}=h^{k\nu}\left(\partial_{\mu}h^i{}_{\nu}-
h^i{}_{\lambda}\Gamma^{\lambda}{}_{\nu\mu}\right).
\end{equation}
So, for $A^{ik}{}_l=h_l{}^{\mu}A^{ik}{}_{\mu}$ we have
\begin{equation}\label{aconnec1}
\begin{array}{l}
\unaa{0}{0}{0}=\frac{1}{2}Q_2,\;
\unaa{1}{1}{0}=\unaa{2}{2}{0}=\unaa{3}{3}{0}=\frac{1}{2}Q_1,\\
\unantiaa{0}{1}{1}=\unantiaa{0}{2}{2}=\unantiaa{0}{3}{3}=%
{\displaystyle \frac{\dot{R}-2R\left(S-\frac{1}{4}Q_1+
\frac{1}{4}Q_3\right)}%
{R}},\\
\unantiaa{1}{2}{2}=\unantiaa{1}{3}{3}=-{\displaystyle
\frac{\sqrt{1-kr^2}}%
{Rr}},\;
\unantiaa{3}{2}{3}={\displaystyle \frac{\cot\th}{Rr}},\\
\unsymaa{0}{1}{1}=\unsymaa{0}{2}{2}=\unsymaa{0}{3}{3}=
-\frac{1}{2}Q_3,
\end{array}
\end{equation}
where a dot means differentiation with respect to time. Taking
into account (\ref{aconnec1}) it is not difficult to find the
curvature tensor $\f{i}{k}{\mu}{\nu}$ defined by
\begin{equation}
\label{ftensor}
\f{i}{k}{\mu}{\nu}=2\partial_{\left[\mu\right.}
A^{ik}{}_{\left.\nu\right]}%
+2A^i{}_{l\left[\nu\right.}A^{lk}{}_{\left.\mu\right]}.
\end{equation}
Nonvanishing components of this tensor are given by three
curvature functions $A$, $B$, $C$:
\begin{equation}\label{ftensor1}
\begin{array}{l}
\f{0}{1}{0}{1}=\f{0}{2}{0}{2}=\f{0}{3}{0}{3}=A-C,\\
\f{1}{0}{1}{0}=\f{2}{0}{2}{0}=\f{3}{0}{3}{0}=A+C,\\
\f{1}{2}{1}{2}=\f{1}{3}{1}{3}=\f{2}{3}{2}{3}=B,
\end{array}
\end{equation}
where
\begin{equation}\label{abc}
\begin{array}{l}
{\displaystyle A=\frac{\left[\dot{R}-2RS_q\right]^{\cdot}}{R}
+\frac{1}{4}Q_3\left(Q_1+Q_2\right)},\\
B={\displaystyle \frac{k+\left[\dot{R}-2RS_q\right]^{2
\vphantom{2^{[}}}}{%
\mathstrut R^2}}-{\displaystyle \frac{1}{4}Q_3^2},\\
{\displaystyle C=\frac{1}{2}\frac{\dot{R}-2RS_q}{R}\left(
Q_1+Q_2\right)
+\frac{1}{2}\frac{\left(RQ_3\right)^{\cdot}}{R}},\\
S_q=S-\frac{1}4{}Q_1^{\vphantom{[^{[}}}+\frac{1}{4}Q_3,
\end{array}
\end{equation}
and $F^{\left[01\right]}{}_{01}=A$,
$F^{\left(01\right)}{}_{01}=-C$, $F^{\left(12\right)}{}_{12}=0$.
In accordance with (\ref{ftensor1}) one obtains nonvanishing
components of the tensors
$F^\mu{}_\nu=\f{\lambda}{\mu}{\lambda}{\nu}$ and
$\tilde{F}^{\mu}{}_{\nu}=\f{\mu}{\lambda}{\nu}{\lambda}$:
\begin{equation}
\begin{array}{l}\label{ftensor2}
F^0{}_0=3\left(A+C\right),\; F^1{}_1=F^2{}_2=F^3{}_3=A+2B-C,\\
\tilde{F}^0{}_0=3\left(A-C\right),\;
\tilde{F}^1{}_1=\tilde{F}^2{}_2=\tilde{F}^3{}_3=A+2B+C.
\end{array}
\end{equation}
All components of the tensor
$V_{\mu\nu}=F^{\lambda}{}_{\lambda\mu\nu}$ are equal to zero. The
scalar curvature
$F=F^{\mu}{}_{\mu}=\tilde{F}^{\mu}{}_{\mu}=6\left(A+B\right)$.
The Bianchy identities for homogeneous isotropic models are
reduced to the following relation
\begin{equation}
\label{bian}
\dot{B}+2H\left(B-A\right)+4AS_q+CQ_3=0,\qquad
\left(H=\frac{\dot{R}}{R}
\right).
\end{equation}

\section{Gravitational Lagrangian and gra\-vi\-ta\-tional field
equa\-tions}

In the frame of MAGT the role of gauge potentials play the tetrad
$h^i{}_{\mu}$, anholonomic connection $A^{ik}{}_\mu$ and
anholonomic metrics \cite{ma1}. The corresponding field strengths
are the tensors $S^i{}_{\mu\nu}$, $\f{i}{k}{\mu}{\nu}$ and
$Q_{ik\mu}$. For mathematical simplification of further analysis
we shall suppose the tetrad to be orthonormalized. Thus the
anholonomic metrics is fixed in the form of Minkowski metrics
$\eta{}_{ij}$. Then the torsion and nonmetricity tensors can be
represented as functions of gauge potentials
\begin{equation}\label{sq}
\begin{array}{l}
S^i{}_{\mu\nu}=\partial_{\left[\nu\right.}h^i{}_{\left.
\mu\right]}-
h_{k\left[\mu\right.}A^{ik}{}_{\left.\nu\right]},\\
Q^{ik}{}_{\mu}=2A^{(ik)}{}_{\mu}.
\end{array}
\end{equation}
The curvature tensor is determined by eqs. (\ref{ftensor}).

The gravitational Lagrangian in MAGT is a function of
$h^i{}_{\mu}$, $S^i{}_{\mu\nu}$, $\f{i}{k}{\mu}{\nu}$ and
$Q^{ik}{}_{\mu}$
\begin{equation}\label{lgmain}
L_G=h{\cal L}_G\left(h^i{}_{\mu},S^i{}_{\mu\nu},
\f{i}{k}{\mu}{\nu},
Q^{ik}{}_{\mu}\right),\; h=\det\left(h^i{}_{\mu}\right),
\end{equation}
where ${\cal L}_G$ is invariant. The gravitational equations can
be derived by variation of the total action integral $I=\int%
d^4x\left(L_m+L_G\right)$ ($L_m$ is Lagrangian of matter) with
respect to $h^i{}_{\mu}$ and $A^{ik}{}_{\mu}$. In connection with
this the obtained system includes two groups of differential
equations --- 16 $h$-eqs. and 64 $A$-eqs. :
\begin{equation}\label{heq}
H_i{}^{\mu}-\nabla_{\nu} \sigma_i{}^{\mu\nu}=t_i{}^{\mu},
\end{equation}
\begin{equation}\label{aeq}
2\nabla_{\nu}\ph_{ik}{}^{\nu\mu}+\sigma_{ik}{}^{\mu}-
2\ph_{il}{}^{\nu\mu}Q^l{}_{k\nu}-2\Omega_{ik}{}^{\mu}
=-J_{ik}{}^{\mu},
\end{equation}
where the following notations are introduced
\[
\begin{array}{l}
{\displaystyle H_i{}^{\mu}=\frac{1}{h}\frac{
\partial L_G}{\partial h^i{}_\mu},\;
\sigma_i{}^{\mu\nu}=\frac{\partial
{\cal L}_G}{\partial S^i{}_{\mu\nu}},\;
\ph_{ik}{}^{\mu\nu}=\frac{\partial {\cal L}_G}{\partial
\f{i}{k}{\mu}{\nu}},}\\
{\displaystyle
\Omega_{ik}{}^\mu=\frac{\partial {\cal L}_G}{
\partial Q^{ik}{}_{\mu}},\;
t_i{}^\mu=-\frac{1}{h}\frac{\delta L_m^{\vphantom{[^{[}}}}{\delta
h^i{}_\mu},\;
J_{ik}{}^\mu=-\frac{1}{h}\frac{\delta L_m}{\delta A^{ik}{}_\mu}},
\end{array}\]
and $\nabla$ denotes here covariant derivative defined by means
of $(-A^{ik}{}_{\mu})$ and \christ in the case of tetrad and
world indices respectively (for example
$\nabla_\nu{}h^i{}_\mu=
\partial_\nu{}h^i{}_\mu A^i{}_{l\nu}h^l{}_\mu-
\mbox{\christ}h^i{}_\lambda$).

The system of 64 $A$-equations (\ref{aeq}) can be divided into
two groups of equations --- 24 $A^{[ik]}{}_{\mu}$-eqs. and 40
$A^{(ik)}{}_{\mu}$-eqs. corresponding to variation with respect
to antisymmetric and symmetric parts of gravitational potentials
$A^{ik}{}_{\mu}$. According to (\ref{conn}) and (\ref{aconnec})
we have
\begin{equation}
\label{decomp}
A^{[ik]}{}_{\mu}=\stackrel{\scriptstyle (R.-C.)}{A}{}^{
[ik]}{}_\mu+
Q_\mu{}^{[ik]},
\end{equation}
where ${\textstyle \stackrel{\scriptstyle
(R.-C.)}{A}{}^{[ik]}{}_\mu}$ as function of tetrad and torsion
has the same form as in the Riemann-Cartan space-time, and the
symmetric part is defined by (\ref{sq}).

In the frame of WCGT the nonmetricity tensor is given by
\begin{equation}\label{wcnonm}
Q^{ik}{}_{\mu}=\eta^{ik}Q_\mu,
\end{equation}
where $Q_\mu$ is the Weyl's vector. By using the gravitational
Lagrangian of WCGT in the form (\ref{lgmain}) we obtain the
gravitational equations of WCGT by variation of the total action
with respect to $h^i{}_\mu$, $A^{[ik]}{}_{\mu}$ and $Q_\mu$. As
result we have $h$-eqs. in the form (\ref{heq}), and
$A^{[ik]}{}_{\mu}$-eqs. and $Q_\mu$-eqs. in the following form
\begin{equation}\label{wcaeq}
2\nabla_{\nu}\ph_{[ik]}{}^{\nu\mu}+\sigma_{[ik]}{}^{\mu}-
2\ph_{[ik]}{}^{\nu\mu}Q_\nu=-J_{ik}{}^{\mu},
\end{equation}
\begin{equation}\label{wcqeq}
2\nabla_{\nu}\ph_{i}{}^{i\nu\mu}+\sigma_{i}{}^{i\mu}-
2{\Omega_i}^{i\mu}=-J_{i}{}^{i\mu}.
\end{equation}
The system of $A^{[ik]}{}_{\mu}$-eqs. (\ref{wcaeq}) is
antisymmetric part of $A$-eqs. (\ref{aeq}) and $Q_\mu$-eqs.
(\ref{wcqeq}) can be obtained by contraction of (\ref{aeq}) with
the tensor $\eta^{ik}$.

The explicit form of eqs (\ref{heq})-(\ref{aeq}) and also
(\ref{wcaeq}) -- (\ref{wcqeq}) depends on the choice of the
gravitational Lagrangian. Because the explicit form of $L_G$ in
MAGT is unknown, we shall use sufficiently general expression
including both a scalar curvature and different invariants
quadratic in the tensors $\f{i}{k}{\mu}{\nu}$, $S^i{}_{\mu\nu}$
and $Q^{ik}{}_\mu$ built by taking into account their symmetry
properties:
\begin{equation}\label{lg}
{\cal L}_G={\cal
L}_F+{\cal  L}_S+{\cal  L}_Q+{\cal  L}_{QS},
\end{equation}
where
\begin{equation}\label{lg1}
\begin{array}{rcl}
{\cal L}_F&=&  f_0
F+F^{\alpha\beta\mu\nu}\left(f_1F_{\alpha\beta\mu\nu}+f_2%
F_{\beta\alpha\mu\nu}+f_3  F_{\alpha\mu\beta\nu}+f_4
F_{\beta\mu\alpha\nu}+ \right.\\
& &\left.+f_5F_{\mu\nu\alpha\beta}\right)+ f_6F^2+
F^{\mu\nu}\left(f_7
F_{\mu\nu}+f_8   F_{\nu\mu}+%
f_9    \tilde{F}_{\mu\nu}+f_{10}
\tilde{F}_{\nu\mu}\right)+\\
& &  +\tilde{F}^{\mu\nu}\left(f_{11}
\tilde{F}_{\mu\nu}+f_{12}\tilde{F}_{\nu\mu}%
\right)+V^{\mu\nu}\left(f_{13}F_{\mu\nu}+f_{14}
\tilde{F}_{\mu\nu}+f_{15}     V_{\mu\nu}\right)\/,\\
{\cal L}_S&=&S^{\alpha\mu\nu}\left(a_1S_{\alpha\mu\nu}+a_2
S_{\nu\mu\alpha}\right)%
+a_3S^\alpha{}_{\mu\alpha}S_\beta{}^{\mu\beta}\/,\\
{\cal L}_Q&=&k_1  Q_{\mu\nu\lambda}Q^{\mu\nu\lambda}+k_2%
Q_{\mu\nu\lambda}Q^{\mu\lambda\nu}+k_3
Q^\mu{}_{\mu\lambda}Q^\nu{}_\nu{}^{\lambda}+\\
&    &     +k_4Q^\mu{}_{\lambda\mu}Q^{\nu\lambda}{}_\nu+k_5
Q^\mu{}_{\mu\lambda}%
Q^{\nu\lambda}{}_\nu\/,\\
{\cal  L}_{QS}&=& m_1Q_{\mu\nu\lambda}S^{\mu\nu\lambda}+m_2
Q^\alpha{}_{\alpha\lambda}%
S_\beta{}^{\beta\lambda}+m_3
Q^\alpha{}_{\lambda\alpha}S_\beta{}^{\beta\lambda}\/.
\end{array}
\end{equation}
The correspondence principle with GR leads to $f_0=(16\pi
G)^{-1}$ ($G$ is Newton's gravitational constant). The
gravitational Lagrangian contains a large number of indefinite
parameters $f_i\;(i=1,\ldots,15)$, $a_l\;(l=1,2,3)$,
$k_p\;(p=1,\ldots,5)$ ³ $m_s\;(s=1,2,3)$. Some restrictions on
these parameters will be found below as result of analysis of
homogeneous isotropic models.

The tensors $H^i{}_\mu$, $\ph_{ik}{}^{\mu\nu}$,
$\sigma_i{}^{\mu\nu}$ and $\Omega_{ik}{}^\mu$ corresponding to
Lagrangian (\ref{lg})--(\ref{lg1}) have the following form:
\begin{equation}
\begin{array}{l}
H_i{}^\mu=-\left(f_0+2f_6F\right)\left(F^\mu{}_i+
\tilde{F}^\mu{}_i\right)
+4f_1F_{klmi}F^{kl\mu m}+4f_2F^{kl\mu m}F_{lkmi}+\\
\mbox{}+4f_3F_{klmi}F^{k[m\mu]l}+2f_4F_{klmi}\left(F^{l[m\mu]k}
+F^{[\mu|kl|m]}
\right)+
4f_5F_{klmi}F^{[\mu m]kl}-\\
\mbox{}-2f_7\left(F^{k\mu}F_{ki}-F^{kl}F^\mu{}_{kli}\right)-
2f_8\left(F^{\mu k}F_{ki}-F^{kl}F^\mu{}_{lki}\right)
-f_9\left(\tilde{F}^{k\mu}F_{ki}+\tilde{F}_{ki}
F^{k\mu}+\right.\\
\mbox{}\left.+F^{k\mu l}{}_iF_{kl}-F^\mu{}_{kli}
\tilde{F}^{kl}\right)
-f_{10}\left(\tilde{F}^{\mu k}F_{ki}+
\tilde{F}_{ki}F^{\mu k}+F^{k\mu l}{}_iF_{lk}-
F^\mu{}_{lki}\tilde{F}^{kl}\right)-\\
\mbox{}-2f_{11}\left(\tilde{F}_{ki}\tilde{F}^{k\mu}+
F^{k\mu l}{}_i\tilde{F}_{kl}\right)-
2f_{12}\left(\tilde{F}_{ki}\tilde{F}^{\mu k}+
F^{k\mu l}{}_i\tilde{F}_{lk}\right)+
f_{13}\left(2V_{ik}F^{[k\mu]}-
F_{ki}V^{k\mu}+\right.\\
\mbox{}\left.+V^{kl}F^\mu{}_{kli}\right)+
f_{14}\left(2V_{ik}\tilde{F}^{[k\mu]}-
\tilde{F}_{ki}V^{k\mu}-V_{kl}F^{k\mu l}{}_i\right)-
4f_{15}V_{ik}V^{\mu k}-4S^k{}_{i\nu}\left(a_1S_k{}^{\mu\nu}-
\right.\\
\mbox{}\left.-a_2S^{\mu\nu}{}_k-
a_3S^\alpha{}_{\alpha}{}^{[\mu}h_k{}^{\nu]}\right)-
2k_1Q^{kl\mu}Q_{kli}-2k_2Q^{k\mu l}Q_{kli}-
2k_3Q^k{}_k{}^\mu Q^l{}_{li}-\\
\mbox{}-2k_4Q^{k\mu}{}_iQ^{lk}{}_l-2k_5Q^k{}_{k(l}
Q^{l\mu}{}_{i)}-m_1
\left(Q_{kli}S^{kl\mu}-
2Q^{k[\mu l]}S_{kli}\right)-
m_2\left(Q^k{}_k{}^lS^\mu{}_{il}+\right.\\
\mbox{}\left.+Q^k{}_{ki}S^l{}_l{}^\mu+Q^k{}_k{}^\mu
S^l{}_{li}\right)-
m_3\left(2Q^{\mu k}{}_{(i}S^m{}_{|m|k)}+
Q^{kl}{}_kS^\mu{}_{il}\right)-h^i{}_\mu{\cal L}_G,
\end{array}
\end{equation}
\begin{equation}
\begin{array}{l}
\sigma_i{}^{\mu\nu}=2\left(a_1S_i{}^{[\mu\nu]}-
a_2S^{\mu\nu}{}_i
-a_3S^\alpha{}_{\alpha}{}^{[\mu}h_i{}^{\nu]}\right)+
m_1Q_i{}^{[\mu\nu]}+\left(m_2
Q^\lambda{}_\lambda{}^{[\nu}+
m_3Q^{\lambda[\nu}{}_{\lambda}\right)h_i{}^{\mu]}
\end{array}
\end{equation}
\begin{equation}
\begin{array}{l}
\ph_{ik}{}^{\mu\nu}=(f_0+2f_6F)h_i{}^{[\mu}h_k{}^{\nu]}+2f_1
F_{ik}{}^{\mu\nu}+2f_2F_{ki}{}^{\mu\nu}+
2f_3F_i{}^{[\mu}{}_k{}^{\nu]}+
f_4\left(F_k{}^{[\mu}{}_i{}^{\nu]}+\right.\\
\mbox{}\left.+F^{[\mu}{}_{ik}{}^{\nu]}\right)+2f_5
F^{\mu\nu}{}_{ik}+2f_7F_k{}^{[\nu}h_i{}^{\mu]}+
2f_8F^{[\nu}{}_kh_i{}^{\mu]}+
f_9\left(\tilde{F}_k{}^{[\nu}h_i{}^{\mu]}+
F_i{}^{[\mu}h_k{}^{\nu]}\right)+\\
\mbox{}+f_{10}\left(\tilde{F}^{[\nu}{}_kh_i{}^{\mu]}+
F^{[\mu}{}_ih_k{}^{\nu]}\right)+
2f_{11}\tilde{F}_i{}^{[\mu}h_k{}^{\nu]}
+2f_{12}\tilde{F}^{[\mu}{}_i%
h_k{}^{\nu]}+f_{13}\left(F^{[\mu\nu]}\eta_{ik}+
h_i{}^{[\mu}V_k{}^{\nu]}
\right)+\\
\mbox{}+f_{14}\left(\tilde{F}^{[\mu\nu]}\eta_{ik}+
V_i{}^{[\mu}h_k{}^{\nu]}
\right)+2f_{15}V^{\mu\nu}\eta_{ik},
\end{array}
\end{equation}
\begin{equation}
\begin{array}{l}
\Omega_{ik}{}^{\mu}=2k_1Q_{ik}{}^{\mu}+2k_2
Q_{(i}{}^\mu{}_{k)}+
2k_3\eta_{ik}Q^{\nu}{}_\nu{}^{\mu}+
2k_4h_{(i}{}^{\mu}Q^\nu{}_{k)\nu}+
k_5\left(\eta_{ik}Q^{\nu\mu}{}_{\nu}+\right.\\
\mbox{}\left.+Q^\nu{}_{\nu(k}h_{i)}{}^{\mu}\right)+
m_1S_{(ik)}{}^{\mu}+
m_2\eta_{ik}S^\nu{}_\nu{}^\mu+
m_3S^\nu{}_{\nu(k}h_{i)}{}^\mu.
\end{array}
\end{equation}

\section{Gravitational equations for homogeneous isotropic models in
MAGT}
Using the gravitational Lagrangian (\ref{lg}) and also
(\ref{conn}) -- (\ref{ftensor2}) we derive the gravitational
equations for homogeneous isotropic models. In accordance with
the structure of the second and third rank tensors in homogeneous
isotropic space we obtain two $h$-equations ($h^{\hat{0}}{}_0$-\,
$h^{\hat{1}}{}_1$-eqs.) and four $A$-equations ($\una{0}{0}{0}$-
\ ,$\una{1}{1}{0}$-\ , $\unsyma{0}{1}{1}$-\ ,
$\unantia{0}{1}{1}$-eqs.). In order to simplify calculations
REDUCE computer algebra system was used. The obtained equations
have the following form:
\begin{equation}
\begin{array}{l}\label{heq1_1}
\tthreefour{0}{0}=6f_0B-12f\left(A^2-B^2\right)-
12C\left(\Bi C+
\Bii A\right)+3aS^2-3\left(k_1+3k_3\right)Q_1^2+\\
\mbox{}+3\left(2k_3+k_5\right)Q_1Q_2-\left(k_1+k_2+k_3+
k_4+k_5\right)Q_2^2
+3\left(2k_1+k_2+3k_4\right)Q_3^2+\\
\mbox{}+
\frac{3}{2}\left[\left(m_1+3m_2\right)Q_1-\left(m_2+m_3
\right)Q_2-\left(m_1-3m_3\right)Q_3-2aS\right]\left(H+
\frac{1}{2}Q_1\right)+\\
\mbox{}+3\left(m_1-3m_3\right)SQ_3,
\end{array}
\end{equation}
\begin{equation}
\begin{array}{l}
\tthreefour{1}{1}=2f_0\left(2A+B\right)+4f\left(A^2-
B^2\right)+4C\left(\Bi
C+\Bii A\right)-3aS^2+3\left(k_1+3k_3\right)Q_1^2-\\
\mbox{}-
3\left(2k_3+k_5\right)Q_1Q_2+2\left(2k_2+3k_5\right)Q_1Q_3+
\left(k_1+k_2+k_3+k_4+k_5\right)Q_{2}^2-\\
\mbox{}-2\left(2k_4+k_5\right)Q_2Q_3
+\left(2k_1+k_2+3k_4\right)Q_3^2+\left[\left(m_1+3m_2
\right)Q_1-
\left(m_2+m_3\right)Q_2-\right.\\
\mbox{}\left.-\left(m_1-3m_3\right)Q_3-2aS\right]
\left(H-2S-\frac{1}{4}Q_1\right)+
\frac{1}{2}\left[\left(m_1+3m_2\right)\dot{Q_1}-
\frac{1}{2}\left(m_2+m_3\right)\dot{Q_2}-\right.\\
\mbox{}\left.-\frac{1}{2}
\left(m_1-3m_3\right)\dot{Q_3}-2a\dot{S}\right],
\end{array}
\end{equation}
\begin{equation}
\begin{array}{l}
\jss{0}{0}{0}=-6\left(\frac{f_0}{2}+2fA-\Biii B+\Bii C
\right)Q_3-
12\left(H-2S_q\right)\left(\Bii \left(A-B\right)+2\Bi C
\right)+\\
\mbox{}+6\left(2k_3+k_5\right)Q_1-4\left(k_1+k_2+k_3+
k_4+k_5\right)Q_2
+6\left(2k_4+k_5\right)Q_3-6\left(m_2+m_3\right)S,
\end{array}
\end{equation}
\begin{equation}
\begin{array}{l}
\jss{0}{1}{1}=-\left(\frac{f_0}{2}+2fA-\Biii B+\Bii
C\right)
\left(Q_1+Q_2\right)+2\left(\frac{f_0}{2}+2fB-\Biii A-\Bii
C\right)Q_3+\\
\mbox{}+
4H\left[\Bii \left(A-B\right)+2\Bi C\right]+
2\left[\Bii \left(\dot{A}-\dot{B}\right)+2\Bi \dot{C}
\right]-
\left(2k_2+3k_5\right)Q_1+\\
\mbox{}+\left(2k_4+k_5\right)Q_2-
2\left(2k_1+k_2+3k_4\right)Q_3-\frac{1}{4}\left[\left(m_1+3m_2
\right)Q_1
-\left(m_2+m_3\right)Q_2-\right.\\
\mbox{}\left.-\left(m_1-3m_3\right)Q_3-2aS\right]-\left(m_1-
3m_3\right)S,
\end{array}
\end{equation}
\begin{equation}
\begin{array}{l}
\jss{1}{1}{0}=2\left(\frac{f_0}{2}+2fA-\Biii B+\Bii
C\right)Q_3+4
\left(H-2S_q\right)\left(\Bii \left(A-B\right)+2\Bi
C\right)-\\
\mbox{}+4\left(k_1+3k_3\right)Q_1-2\left(2k_3+k_5
\right)Q_2+
2\left(2k_2+3k_5\right)Q_3-\frac{1}{2}\left[\left(m_1+
3m_2\right)Q_1-\right.\\
\mbox{}\left.-\left(m_2+m_3\right)Q_2-\left(m_1-3m_3
\right)Q_3
-2aS\right]-2\left(m_1+3m_2\right)S,
\end{array}
\end{equation}
\begin{equation}\label{aeq1_4}
\begin{array}{l}
\jaa{0}{1}{1}=16f\left(\dot{A}+\dot{B}\right)+
16\left[f_0+
4f\left(A+B\right)\right]S_q+8\left(\Biii
+2f\right)CQ_3-\\
\mbox{}-\left[\left(m_1+3m_2\right)Q_1-\left(m_2+
m_3\right)Q_2-\left(m_1-3m_3\right)Q_3-2aS\right]+\\
\mbox{}+8\Bii\dot{C}-4\left[2\Bi C+\Bii\left(A-B
\right)\right]
\left(Q_1+Q_2\right)+32\Bii \left(H-S_q\right)C,
\end{array}
\end{equation}
where the following notations are introduced:
\begin{equation}
\displaystyle
\begin{array}{l}
a=2a_1+a_2+3a_3,\\
f=f_1-f_2+\frac{f_3-f_4}{2}+f_5+3f_6+
\sum\limits_{i=7}^{12}f_i,\\
\Bi=f_1+f_2+\frac{f_3+f_4}{2}+f_7+f_8-f_9-f_{10}
+f_{11}+f_{12},\\
\Bii=f_7+f_8-f_{11}-f_{12},\\
\Biii=2\Bi-\Bii-2f-4f_2-2f_4+2f_5+3f_9+3f_{10}
-2f_{11}-2f_{12}=\\
\phantom{\Biii}=-6f_6-f_7-f_8-f_9-f_{10}
-f_{11}-f_{12}.
\end{array}
\end{equation}

The system of gravitational equations (\ref{heq1_1}) --
(\ref{aeq1_4}) includes 12 coefficients $f_i$ ($i=1,\ldots ,12$)
in the form of four combinations $f$, $\Bi$, $\Bii$, $\Biii$.
Three coefficients $a_k$ ($k=1,2,3$) appear in equations by means
of $a$. Coefficients $k_i$ and $m_s$ appear in various
combinations.

Further analysis of obtained system of equations is connected
with searching of such restrictions on indefinite parameters,
which essentially simplify investigated system and leads to
cosmological equations, which are reasonable generalization of
the cosmological Friedmann equations of GR.

As source of gravitational field we will consider fluid (field)
with the following nonvanishing components of the energy-momentum
tensor $t_{\hat{0}}{}^{\hat{0}}=\rho$, $t_{\hat{1}}{}^{\hat{1}}=
t_{\hat{2}}{}^{\hat{2}}=t_{\hat{3}}{}^{\hat{3}}=-p$ ($\rho$ is
the energy (mass) density, $p$ is the pressure) and with
vanishing tensor $J_{ik}{}^{\mu}=0$.

\section{Generalized cos\-mo\-lo\-gi\-cal Fried\-mann equa\-tion and
homogeneous isotropic models in the Weyl-Cartan space-time}

At first, let us put $k_i=0$ and $m_k=0$. It means that terms of
$L_G$ including nonmetricity are omitted. Then the gravitational
equations (\ref{heq1_1})--(\ref{aeq1_4}) can be transformed to
the form
\begin{equation}\label{heq3_1}
\begin{array}{l}
6f_0B-12f\left(A^2-B^2\right)-12C\left(\Bi C+\Bii
A\right)+3aS^2
-\frac{3}{2}a\left(Q_1+2H\right)S=\rho,
\end{array}
\end{equation}
\begin{equation}\label{heq3_2}
\begin{array}{l}
2f_0\left(2A+B\right)+4f\left(A^2-B^2\right)+
4C\left(\Bi C +\Bii A\right)
+aS^2-a\dot{S}+\\
\mbox{}+\frac{1}{2}a\left(Q_1-4H\right)S=-p,
\end{array}
\end{equation}
\begin{equation}
\label{aeq3_1}
\begin{array}{l}
\left[f_0+2\Bii C-2\Biii B+4fA\right]Q_3+
4\left(H-2S_q\right)
\left[2\Bi C+\Bii \left(A-B\right)\right]=0,
\end{array}
\end{equation}
\begin{equation}
\begin{array}{l}
\label{aeq3_2}
\left[\frac{f_0}{2}+2fA-\Biii B+\Bii C\right]
\left(Q_1+Q_2\right)-2\left(\frac{f_0}{2}+2fB-\Biii A-
\Bii C\right)Q_3
-\\
\mbox{}-4\left(\Bii \left(A-B\right)+2\Bi C\right)H-
2\left(\Bii \left(\dot{A}-\dot{B}\right)+2\Bi
\dot{C}\right)-\frac{1}{2}aS=0,
\end{array}
\end{equation}
\begin{equation}
\label{aeq3_3}
aS=0
\end{equation}
\begin{equation}
\label{aeq3_4}
\begin{array}{l}
4f\left(\dot{A}+\dot{B}\right)+
4\left[f_0+4f\left(A+B\right)\right]S_q
+2\left(2f+\Biii\right)CQ_3 +\frac{1}{2}aS-\\
\mbox{}-\left[\Bii\left(A-B\right)+2\Bi C
\right]\left(Q_1+Q_2\right) +2\Bii
\left[\dot{C}+4\left(H-S_q\right)C\right]=0
\end{array}
\end{equation}
If one imposes on the parameters two following conditions
\begin{equation}
\label{cond1}
\Bi=0,\qquad \Bii=0,
\end{equation}
then equations (\ref{aeq3_1}) -- (\ref{aeq3_3}) give
\begin{equation}\label{cons}
Q_1+Q_2=0,\qquad Q_3=0.
\end{equation}
By means of (\ref{cons}) and (\ref{cond1}) equations
(\ref{heq3_1}), (\ref{heq3_2}) and (\ref{aeq3_4}) take the form
\begin{equation}
\left.\label{syst1}
\begin{array}{l}
6f_0B-12f\left(A^2-B^2\right)=\rho\\
2f_0\left(2A+B\right)+4f\left(A^2-B^2\right)=-p\\
4f\left(\dot{A}+\dot{B}\right)+4\left[f_0+4f
\left(A+B\right)\right]S_q=0\\
\end{array}\right\}
\end{equation}
Equations (\ref{syst1}) lead to the following solution
\begin{equation}
\label{oldsol}\left\{
\begin{array}{l}
S-{\displaystyle \frac{1}{4}Q_1}={\displaystyle
-\frac{1}{4}}
{\displaystyle \frac{d}{dt}}\ln{\left|1
-\beta\left(\rho-3p\right)\right|}\\
A=-{\displaystyle \frac{1}{12f_0}}{\displaystyle
\frac{\rho+3p+\frac{\beta}{2}\left(\rho-3p\right)^2}%
{1-\beta\left(\rho-3p\right)}}\\
B={\displaystyle \frac{1}{6f_0}}
{\displaystyle \frac{\rho-\frac{\beta}{4}\left(\rho
-3p\right)^2}{1-\beta\left(\rho-3p\right)}}\\
\end{array}\right.
\end{equation}
where $\displaystyle \beta=-\frac{f}{3f_0\,^2}$. System of
equation (\ref{syst1}) and solution (\ref{oldsol}) was found
earlier in ref. \cite{ma4}, where the function $Q_3$ was not
taken into consideration.

As a consequence of eq. (\ref{aeq3_3}) the torsion $S$ is
determined and equals to zero, if $a\neq 0$. It is easy to verify
that solution (\ref{oldsol}) takes also place, if terms with
coefficients $m_i$ ($i = 1,2,3$) in ${\cal L}_G$ are taken into
account and the following two conditions are imposed
\begin{equation}\label{cond1_1}
m_1+3m_2=0,\quad m_1-3m_3=0.
\end{equation}
Note, that solution (\ref{oldsol}) takes place in WCGT by the
same restrictions on indefinite parameters of $L_G$.

From geometric point of view, obtained solution (\ref{oldsol}),
(\ref{cons}) corresponds to the Weyl space-time and the
nonmetricity is determined by the only function $Q_1$, which is
the time component of the Weyl's vector. The solution
(\ref{oldsol}) leads to the generalized cosmological Friedmann
equation (GCFE) for the scale factor
\begin{equation}\label{gcfe}
\frac{k}{R^2}+\left\{
\frac{d}{dt}\ln\left[R\sqrt{\left|1-\beta(\rho
-3p)\right|\,}\,\right]\right\}^2=
{\displaystyle \frac{1}{6f_0}\frac{\rho-\frac{\beta}{4}
\left(\rho-3p\right)^2}{1-\beta\left(\rho-3p\right)}}.
\end{equation}
GCFE (\ref{gcfe}) can be derived by using expressions (\ref{abc})
and (\ref{oldsol}) for the function $B$. At first, GCFE
(\ref{gcfe}) was obtained in the frame of PGT \cite{ma3}. Its
validity in MAGT based on the gravitational Lagrangian with
$k_i=0$ and $m_s=0$ was proved in \cite{ma4}. It is important,
that GCFE is a first order differential equation like the
Friedmann cosmological equation of GR. GCFE satisfies the
correspondence principle with GR in the case of sufficiently
small energy densities $\rho\ll|\beta|^{-1}$, under which
non-einsteinian characteristics (torsion and nonmetricity) are
negligible.

Now let us investigate the system of gravitational equations of
MAGT (\ref{heq1_1})--(\ref{aeq1_4}) for homogeneous isotropic
models in the Weyl-Cartan space-time by using general
gravitational Lagrangian (\ref{lg})--(\ref{lg1}). Then
nonmetricity functions satisfy relations (\ref{cons}), and
conditions (\ref{cond1}) are assumed to be fulfilled. Equations
(\ref{heq1_1}) -- (\ref{aeq1_4}) take the form
\begin{equation}\label{heq4_1}
\begin{array}{l}
6f_0B-12f\left(A^2-B^2\right)+3aS^2-\left(k-
\frac{1}{4}m\right)Q_1^2
-\frac{3}{2}a\left(Q_1+2H\right)S+\frac{3}{2}mQ_1H=\rho,
\end{array}
\end{equation}
\begin{equation}\label{heq4_2}
\begin{array}{l}
2f_0\left(2A+B\right)+4f\left(A^2-B^2\right)+aS^2+\left(k-
\frac{1}{4}m\right)Q_1^2-2\left(m-
\frac{1}{4}a\right)Q_1S+\\
\mbox{}+\left(mQ_1-2aS\right)H+
\frac{1}{2}\left(m\dot{Q}_1-2a\dot{S}\right)=-p,
\end{array}
\end{equation}
\begin{equation}\label{aeq4_1}
\left(2k_1+2k_2+8k_3+2k_4+5k_5\right)Q_1-3\left(m_2+
m_3\right)S=0,
\end{equation}
\begin{equation}\label{aeq4_2}
2\left(k_2+k_4+2k_5+\frac{1}{8}m\right)Q_1+\left(m_1-3m_3-
\frac{1}{2}a\right)S=0,
\end{equation}
\begin{equation}\label{aeq4_3}
\left(4k-\frac{3}{2}m\right)Q_1-3\left(2m-a\right)S=0,
\end{equation}
\begin{equation}
\label{aeq4_4}
f\left(\dot{A}+\dot{B}\right)+\left[f_0+4f\left(A+B
\right)\right]S_q+\frac{1}{4}\left(2aS-mQ_1\right)=0,
\end{equation}
where the following notations are introduced
\[
\begin{array}{rcl}
k&=&4k_1+k_2+16k_3+k_4+4k_5,\\
m&=&m_1+4m_2+m_3.
\end{array}
\]

Note that only two equations from three
(\ref{aeq4_1})--(\ref{aeq4_3}) are linearly independent,
therefore eq. (\ref{aeq4_1}) will be excluded from further
consideration. The analysis of gravitational equations
(\ref{heq4_1}), (\ref{heq4_2}), (\ref{aeq4_2})--(\ref{aeq4_4})
leads us to the solution (\ref{oldsol}) corresponding to
different models in the Riemann-Cartan, Weyl, Weyl-Cartan
space-times in dependence on restrictions on parameters of $L_G$.
\begin{enumerate}
\item If $a=0$, $m=0$, $m_1-3m_3=0$, and $k\neq 0$ and/or
$k_2+k_4+2k_5\neq 0$, we have $Q_1=0$ that corresponds to models
in the Riemann-Cartan space-time considered at the first time in
the frame of PGT \cite{ma3}.
\item If $k=m=k_2+k_4+2k_5=0$, and $a\neq 0$ and/or
$m_1-3m_3-\frac{a}{2}\neq 0$, we have $S=0$ that corresponds to
models in the Weyl space-time.
\item Isotropic models in the Weyl-Cartan space-time ($S\neq 0$,
$Q_1\neq 0$) take place, if $a\neq 0$, $m\neq 0$ and $3m^2=4ak$
and, in addition, other restrictions on indefinite parameters are
fulfilled:
\begin{enumerate}
\item $k_2+k_4+2k_5+\frac{m}{8}=0$, $m_1-3m_3-\frac{a}{2}=0$,
$2m-a\neq 0$ $8k-3m\neq 0$ or
\item $3m(m_2+m_3)=2a(2k_1+2k_2+8k_3+2k_4+5k_5)$, $8k-3m\neq 0$,
$2m-a\neq 0$, $k_2+k_4+2k_5+\frac{m}{8}\neq 0$,
$m_1-3m_3-\frac{a}{2}\neq 0$.
\end{enumerate}
According to (\ref{oldsol})  and (\ref{aeq4_3})
\begin{equation}\label{tornonm1}
\left\{
\begin{array}{l}
Q_1={\displaystyle \frac{a}{a-2m}}\:
{\displaystyle \frac{d}{dt}}\ln{\left|1-
\beta\left(\rho-3p\right)%
\right|},\\
S={\displaystyle \frac{m}{2(a-2m)}}\:
{\displaystyle \frac{d}{dt}}\ln{\left|1-
\beta\left(\rho-3p\right)\right|}.\\
\end{array}\right.
\end{equation}
\end{enumerate}
Besides indicated two possibilities, solution for isotropic
models in the Weyl-Cartan space-time can be obtained, if
$k=a=m=0$, $k_2+k_4+2k_5\neq 0$, $m_1-3m_3\neq 0$,
$m_1-3m_3+8(k_2+k_4+2k_5)\neq 0$, then
\[
Q_1=-\frac{m_1-3m_3}{2(k_2+k_4+2k_5)}S
\]
and according to (\ref{oldsol})
\[\left\{
\begin{array}{l}
Q_1={\displaystyle \frac{m_1-3m_3}{8(k_2+k_4+2k_5)+
m_1-3m_3}}\:
{\displaystyle \frac{d}{dt}}\ln{\left|1-
\beta\left(\rho-3p\right)\right|},\\
S={\displaystyle -\frac{2(k_2+k_4+2k_5)}{8(k_2+
k_4+2k_5)+m_1-3m_3}}\:
{\displaystyle \frac{d}{dt}}\ln{\left|1-
\beta\left(\rho-3p\right)\right|}.\\
\end{array}\right.\]

In the frame of WCGT the system of gravitational equations for
homogeneous isotropic models is reduced to (\ref{heq4_1}),
(\ref{heq4_2}), (\ref{aeq4_3}) and (\ref{aeq4_4}). The solution
(\ref{oldsol}) corresponds to different models in dependence on
restrictions on parameters of $L_G$:
\begin{enumerate}
\item If $m=a=0$ and $k\neq 0$ we have models in the
Riemann-Cartan space-time with $S\neq 0$ and $Q_1=0$.
\item If $m=k=0$ and $a\neq 0$ we have $S=0$ and $Q_1\neq 0$ that
corresponds to models in in the Weyl space-time.
\item If $3m^2-4ak=0$ and $k-\frac{3}{8}m\neq 0$, $2m-a\neq 0$ we
have the models in the Weyl-Cartan space-time and the torsion and
nonmetricity are defined by (\ref{tornonm1}).
\end{enumerate}

Note, that the metrics of all models indicated above satisfies
the GCFE (\ref{gcfe}). This means that various regular
cosmological solutions of GCFE obtained earlier in the frame of
PGT (see refs. in \cite{ma2}) take place also in the frame of
MAGT (WCGT). Hence, the most important physical consequences
obtained in the frame of PGT --- the possible existence of
limiting energy density for usual gravitating systems, the
repulsion gravitational effect provoked by various physical
factors etc. --- are valid also in MAGT (WCGT).

\end{document}